\documentclass[a4paper,fleqn]{cas-dc}

\usepackage[numbers]{natbib} 
\usepackage[utf8]{inputenc}

\def\tsc#1{\csdef{#1}{\textsc{\lowercase{#1}}\xspace}}
\tsc{WGM}
\tsc{QE}
\tsc{EP}
\tsc{PMS}
\tsc{BEC}
\tsc{DE}

\begin{document}

\let\WriteBookmarks\relax
\def\floatpagepagefraction{1}
\def\textpagefraction{.001}
\shorttitle{H$_{2}$-diluted precursors for GaAs doping in chemical beam epitaxy}
\shortauthors{K. Ben Saddik et~al.}

\title [mode = title]{H$_{2}$-diluted precursors for GaAs doping in chemical beam epitaxy}                      



\author{K. Ben~Saddik}[type=editor,
                        auid=000,bioid=1,
                        orcid=0000-0003-0060-5651]
\cormark[1]
\ead{karim.ben@uam.es}


\address{Electronics and Semiconductors Group (ElySe), Applied Physics Department, Universidad Autónoma de Madrid,
ES-28049 Madrid, Spain}

\author%
{A. F. Braña}[type=editor,
                        auid=000,bioid=1,
                        orcid=0000-0002-2125-8495]
\ead{alejandro.brana@uam.es}

\author{N. López}[type=editor,
                        auid=000,bioid=1,
                        orcid=0000-0001-6510-1329]
\ead{n.lopez@uam.es}

\author%
{B. J. García}[type=editor,
                        auid=000,bioid=1,
                        orcid=0000-0002-6711-2352]
\cormark[2]
\ead{basilio.javier.garcia@uam.es}

\author{S. Fernández-Garrido}[type=editor,
                        auid=000,bioid=1,
                        orcid=0000-0002-1246-6073]
\cormark[2]
\ead{sergio.fernandezg@uam.es}


\cortext[cor1]{Corresponding author}
\cortext[cor2]{Principal corresponding authors}
\hyphenation{hol-der}
\hyphenation{tem-pe-ra-tu-re}
\hyphenation{ba-sed}
\hyphenation{so-ur-ces}
\hyphenation{non-in-ten-tio-na-lly}

\begin{abstract}
A wide range of n- and p-type doping levels in GaAs layers grown by chemical beam epitaxy is achieved using H$_{2}$-diluted DTBSi and CBr$_{4}$ as gas precursors for Si and C. We show that the doping level can be varied by modifying either the concentration or the flux of the diluted precursor. Specifically, we demonstrate carrier concentrations of $6\times10^{17}$--$1.2\times10^{19}$~cm$^{-3}$ for Si, and $9\times10^{16}$--$3.7\times10^{20}$~cm$^{-3}$ for C, as determined by Hall effect measurements. The incorporation of Si and C as a function of the flux of the corresponding diluted precursor is found to follow, respectively, a first and a fourth order power law. The dependence of the electron and hole mobility values on the carrier concentration as well as the analysis of the layers by low-temperature (12~K) photoluminescence spectroscopy indicate that the use of H$_{2}$ for diluting DTBSi or CBr$_{4}$ has no effect on the electrical and optical properties of GaAs.
\end{abstract}


 

\begin{keywords}
A1. Doping \sep A3. Chemical beam epitaxy \sep  B1. Inorganic compounds \sep B2. Semiconducting III-V materials \sep B2. Semiconducting gallium arsenide \end{keywords}.

\maketitle

\section{Introduction}

Solid-state opto- and electronic devices are typically based on junctions formed by the combination of semiconducting layers with varying electrical conductivities, i.\,e., with different doping characteristics. The accurate control of the dopant concentration and the possibility of tuning it over a wide range are both essential requirements for the fabrication of certain types of devices, such as heterojunction bipolar transistors and multi-junction solar cells. In the particular case of III-As compound semiconductors, two of the most commonly used elements for n- and p-type doping are Si and C, respectively. These two elements are quite often preferred over other alternatives because of both their comparatively low diffusion coefficients, which facilitate the fabrication of junctions with abrupt doping profiles \citep{Furuhata1988,Cunningham1989}, and their high solubility limits \cite{Madelung1996,Newman1999}. 

In chemical beam epitaxy (CBE), where gas sources - such as metalorganic compounds and hydrides - are used as precursors \cite{Tsang1991a}, silane and disilane are the typical gases employed for Si doping. The use of these precursors makes it possible to achieve electron concentrations in GaAs of up to 1--$5\times10^{18}$~cm$^{-3}$ \citep{Kimura1987,Heinecke1987}. These values are close to the Si solubility limit at the typical substrate temperatures used in CBE for the growth of this compound ($500$--$600~^{\circ}$C), particulary, about $4\times10^{18}$~cm$^{-3}$ \cite{Madelung1996,Newman1999}. Both silane and disilane are, however, highly hazardous gases with very high vapor pressures at room temperature. Specifically, they are toxic as well as extremely flammable due to their pyrophoric character. In this context, ditertiarybutylsilane (DTBSi) is an interesting alternative for Si doping since it is a non-toxic and non-pyrophoric liquid that can also be properly refined to meet the purity standards of the semiconductor industry \cite{Leu1998,Fulem2005}. In addition, at room temperature, DTBSi has a subatmospheric  vapor pressure, which is high enough for the direct doping of III-V compounds \cite{Fulem2005}. Surprisingly, even though the first study on the use of this precursor in metal-organic chemical vapor epitaxy (MOVPE) dates from the late nineties~\cite{Leu1998}, we are not aware of any report on the n-type doping of GaAs layers using DTBSi in CBE. With respect to C doping, which has a solubility limit in GaAs of approximately $1\times10^{19}$~cm$^{-3}$ \cite{Madelung1996,Newman1999}, carbon tetrabromide CBr$_{4}$ is one of the common precursors in CBE as well as in gas- and solid-source molecular beam epitaxy (MBE) \cite{Lyon1991,Zhang1993,Huang2003,Chang2010}. This precursor provides very high hole concentration in GaAs, on the order of $3\times10^{19}$--$1\times10^{20}$~cm$^{-3}$ \cite{Lyon1991,Houng1993,Westwater1997}. Nevertheless, the real challenge when using CBr$_{4}$ is to obtain, not high, but low doping levels. So far, two different approaches were reported to achieve moderate C concentrations using this precursor. The most extended one consists in decreasing the CBr$_{4}$ flux by lowering the temperature of the liquid reservoir, therefore, decreasing its vapor pressure \cite{Huang2003}. The second approach is based on controlling the CBr$_{4}$ flux by varying the conductance of the gas line in its way from the reservoir to the growth chamber \cite{Chang2010}. These approaches yielded hole concentrations down to $1\times10^{18}$~cm$^{-3}$ \cite{Huang2003,Chang2010}. However, this value is still too high for the fabrication of certain devices, such as GaAs based solar cells, which demand doping concentrations on the order of $1\times10^{17}$~cm$^{-3}$ \cite{Kurtz1999,Friedman1998}. Hence, there is a clear need for the development of more suitable solutions.

In this work, we analyze the possibility of controlling the Si and C doping levels in CBE-grown GaAs layers by using controlled dilutions of DTBSi or CBr$_{4}$ in H$_{2}$. H$_{2}$ is thus not intended to be used as a simple carrier gas using a bubbler, but to control the actual flux of the dopant precursor. Such an approach is based on previous studies about the growth of GaAs layers doped with Sn \cite{Nunez2011} and on the use of H$_{2}$-diluted precursors in chemical vapor deposition (CVD) to decrease the risk associated to the use of hazardous gases as well as to improve the performance of amorphous-Si solar cells \cite{Tsu1997,Yue2001}. The analysis of the doping concentration by Hall effect measurements as a function of both the concentration and the flux of the H$_{2}$-diluted precursor demonstrates that, by using this approach, it is possible to modify the Si and C concentrations over at least two and three decades, respectively. Importantly, we also find that the use of H$_{2}$ to dilute the precursors has no detrimental effects on either the Hall mobility or the luminescence properties, as shown in the latter case by low-temperature photoluminescence (PL) spectroscopy.

\section{Experiment}
Epi-ready and semi-insulating GaAs(100) wafers, purchased from Wafer Technology LTD, are used as substrates for the growth of Si- and C-doped GaAs layers. The carrier concentration of the wafers is $7.3\times10^{6}$--$2.8\times10^{7}$~cm$^{-3}$. For our experiments, the as received substrates are In-bonded onto a Mo holder before being loaded into a Riber CBE32 system. A detailed description of the CBE system can be found elsewhere \cite{Ait-Lhouss1994}. 
In order to desorb the protective oxide layer, the substrates are first outgassed at $580~^{\circ}$C, as measured by an infrared optical pyrometer, in vacuum until observing the characteristic reflection high-energy electron di-ffraction (RHEED) pattern of GaAs. Subsequently, the substrates are further outgassed for about 5~min at $600~^{\circ}$C under the As precursor flux, which results in the observation of a faint ($2\times4$) surface reconstruction. Finally, the substrate temperature is decreased down to $530~^{\circ}$C for the growth of either Si- or C-doped GaAs. The growth rate of the GaAs doped layers is set to 0.5~$\mu$m/h, as measured by RHEED intensity oscillations frequency, to obtain final thicknesses ranging from 0.5 to 1~$\mu$m/h. As gaseous precursors for Ga, As, Si and C, we use triethylgallium (TEGa), tertiarybutylarsine (TBAs), DTBSi, and CBr$_{4}$, respectively. Low-temperature (120$~^{\circ}$C) gas injectors are used for TEGa and CBr$_{4}$, while high-temperature  ones are employed for TBAs ($820~^{\circ}$C) and DTBSi ($1000~^{\circ}$C) to achieve a high cracking efficiency, as observed by quadrupole mass spectrometry. H$_{2}$-diluted precursors are prepared using 99.999$\%$ pure H$_{2}$, which is further purified by a heated Ag-Pd membrane. The dilution of DTBSi or CBr$_{4}$ is  carried out inside a non-heated intermediate reservoir, which is used as dopant source during growth. This reservoir is initially filled with either DTBSi or CBr$_{4}$ and then with H$_{2}$ until reaching the desired dopant precursor concentration, hereafter referred as dilution factor, which is defined as:
\begin{equation}
\label{concentration}
f=\frac{P_{dopant}}{P_{dopant}+P_{H_{_{2}}}},
\end{equation}
where \textit{P}$_{H_{2}}$ and \textit{P$_{dopant}$} are, respectively, the H$_{2}$ and the dopant partial pressures inside the reservoir. These partial pressures are measured using a capacitance manometer, whose response is almost independent of the gas type. Precursor fluxes are controlled by a pressure-regulated control system that determines the flux through a hole at the exit of an intermediate chamber located between the precursor container and the growth chamber. The arrival rates of the H$_{2}$-diluted precursors to the substrate surface are given here as beam equivalent pressures (BEP). The BEP values are measured before the growth using an ion gauge, which is placed at the same position as the substrate during growth.

Hall effect measurements are performed at room temperature using In contacts and $\thickapprox$ 1~cm$^{2}$ van der Pauw patterns. For the analysis of the Hall data, we take into account the thickness of the surface depletion layer associated to the Fermi level pinning at the GaAs surface. According to Ref.~\cite{Holloway1995}, we assume that for n-type GaAs the Fermi level is pinned at 0.68~eV below the conduction band minimum, and for p-type 0.5~eV above the valence band maximum. The analysis of the samples by photoluminescence (PL) spectroscopy is carried out at 12~K using a closed-cycle He cryostat. The PL is excited with an Ar laser, dispersed by an 50-cm Spex monochromator, and detected by an a combined Si/(In,Ga)As photodiode. The excitation power density is about 1~kW/cm$^{2}$. 

\section{Results and discussion}

\subsection{GaAs doping using H$_{2}$-diluted DTBSi}

\begin{figure*}
\includegraphics*[width=1\textwidth]{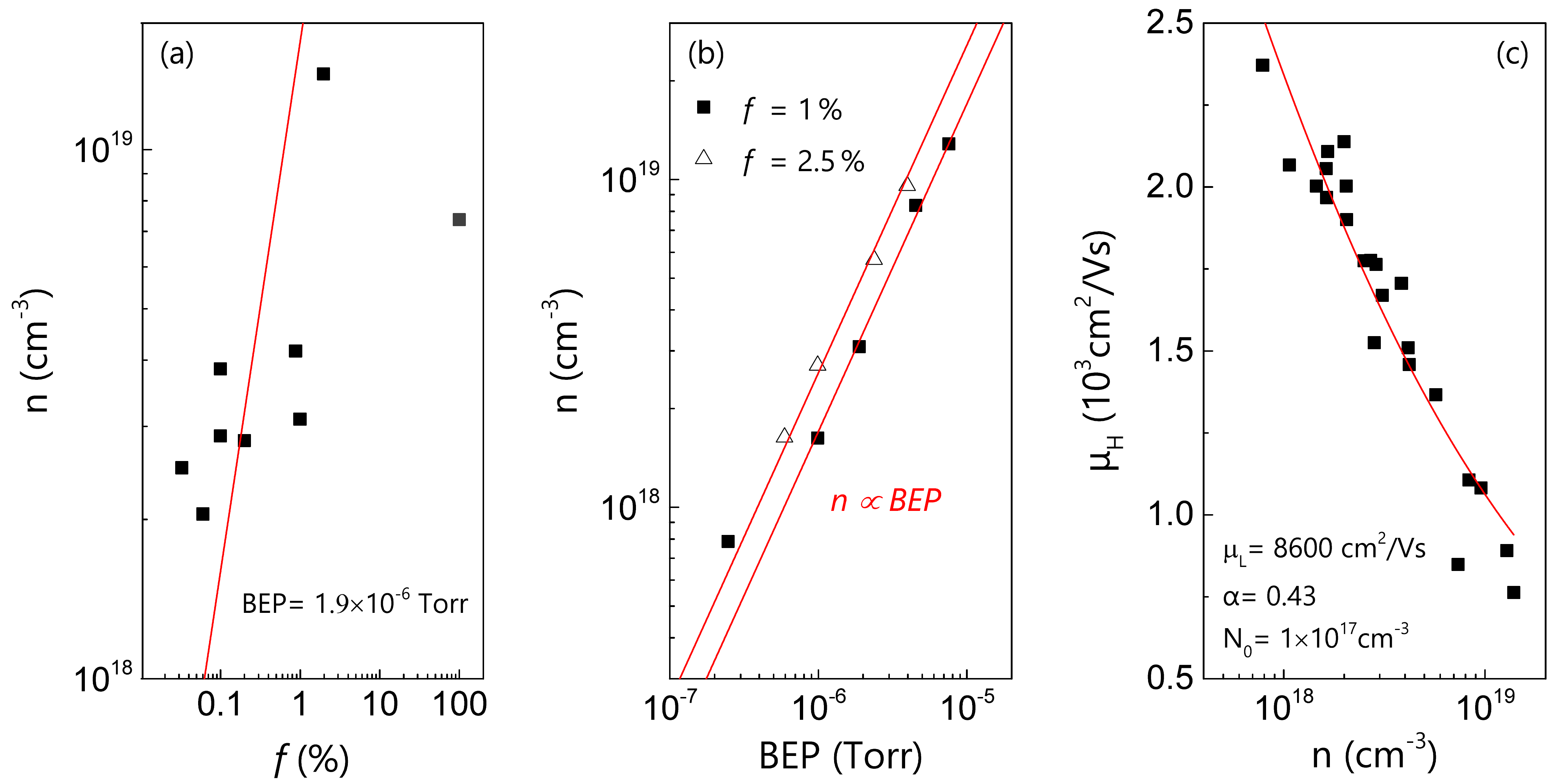}
\caption{\label{DTBSi} Electron concentration, as determined by Hall effect measurements, as a function of (a) the dilution factor and (b) the BEP of the H$_{2}$-diluted DTBSi precursor. In (a) the BEP is $1.9\times10^{-6}$~Torr and in (b) the dilution factor is either 1 or 2.5\%, as indicated in the legend. The solid lines in (a) and (b) are fits of a first order power law to the experimental data. In (a) the fit is performed for \textit{f} values lower than $2.5\%$. (c) Hall electron mobility dependence on the electron concentration. The solid line is a fit of eq.(\ref{Hilsum}) to the experimental data.}
\end{figure*}

The electron concentration, \textit{n}, in GaAs layers doped with H$_{2}$-diluted DTBSi is investigated first as a function of the DTBSi dilution factor, \textit{f}, as defined by eq.(\ref{concentration}). Figure~\ref{DTBSi}(a) shows the dependence of the Hall electron concentration on \textit{f} for a H$_{2}$-diluted DTBSi BEP value of 1.9$\times10^{-6}$~Torr. The value of \textit{n} increases from about 2$\times10^{18}$ to 1.4$\times10^{19}$~cm$^{-3}$ as \textit{f} is varied between 0.03 and 2~$\%$. The maximum doping concentration achieved here, 1.4 $\times10^{19}$~cm$^{-3}$, is even higher  than the highest values reported in the literature for Si doped GaAs layers using DTBSi as gas precursor \cite{Leu1998}. When using pure DTBSi (100~$\%$), the electron concentration decreases down to 7.4$\times10^{18}$~cm$^{-3}$. We attribute this drop to the self-compensation caused by the formation of Si$_{Ga}$V$_{Ga}$ complexes \cite{Mohades-Kassai2000} or even Si$_{As}$ due to its amphoteric character \cite{Furuhata1988}. Finally, as pointed out in the experimental section, we stress here the necessity of using an elevated temperature for the DTBSi cell in order to achieve a high doping efficiency. Further experiments (not shown here) demonstrate a negligible incorporation of Si as the cell is kept at low temperature, 120~$^{\circ}$C. 

Our results demonstrate that the n-type doping level can be modified by varying the dilution factor of the DTBSi gaseous precursor. However, the observed electron concentration spreading data between 0.03 and 1~$\%$ reveals a poor control on the actual doping level. We attribute the data dispersion to deviations of the actual values of \textit{f} with respect to the nominal ones. Such deviations are likely to occur due to an uncontrolled partial condensation of DTBSi inside the reservoir during the preparation of the gas mixture. This undesirable effect could be avoided by heating the reservoir above the DTBSi condensation point temperature during the dilution process. 

Figure~\ref{DTBSi}(b) presents the variation of the electron concentration with the BEP of the H$_{2}$-diluted DTBSi precursor for two different values of \textit{f}, 1 and 2.5~$\%$. Since for each series of samples we always use the same gas mixture, these data are not distorted by the possible condensation of DTBSi inside the reservoir during its dilution in H$_{2}$. Regardless of the value of \textit{f}, we observe a monotonic increase of \textit{n} with the BEP. The results shown in Fig.~\ref{DTBSi}(b) demonstrate that, for a given value of \textit{f}, the final doping level can be accurately controlled by varying the flux of the H$_{2}$-diluted DTBSi precursor. Specifically, the doping level can be tuned at least from 8$\times10^{17}$ to 1.3$\times10^{19}$~cm$^{-3}$. Interestingly, for the two different values of \textit{f} used in this work, the electron concentration increases linearly with the flux of H$_{2}$-diluted DTBSi, as indicated by the corresponding fits shown in Fig.~\ref{DTBSi}(b). Therefore, the incorporation of Si into GaAs as a function of the BEP follows a first order power law. The same dependence of \textit{n} but with respect to the partial pressure of the precursor was reported in Ref.~\cite{Leu1998} for the MOVPE of GaAs layers doped with pure DTBSi. Such a trend has also been observed in the case of GaN layers doped with DTBSi in MOVPE and CVD \cite{Deatcher2004,Fong2007}. 

Figure~\ref{DTBSi}(c) shows the dependence of the room-temperature Hall electron mobility on the electron concentration for GaAs doped with H$_{2}$-diluted DTBSi. As the electron concentration increases from 8$\times10^{17}$ to 1.4$\times10^{19}$~cm$^{3}$, the mobility decreases from 2.4$\times10^{3}$ to 0.76$\times10^{3}$~cm$^{2}$/V$\cdot$s. These values are comparable to those reported in Refs.~\cite{Furuhata1988,Leu1998,Druminski1982,Kamp1994} for GaAs layers doped with Si using different precursors including DTBSi. We thus conclude that the dilution of DTBSi in H$_{2}$ for the n-type doping of GaAs has no impact on the electron mobility. The dependence of the mobility $\mu$ on the electron concentration is well described by the empirical equation proposed by Hilsum \cite{Hilsum1974}:
\begin{equation}
\label{Hilsum}
\mu=\frac{\mu_{L}}{1+\mu_{L}/\mu_{I}} = \frac{\mu_{L}}{1+\left(N/N_{0}\right)^{\alpha}},
\end{equation}
where $\mu_{L}$ is the lattice scattering mobility, $\mu_{I}$ the impurity scattering mobility, \textit{N} the impurity concentration (in this case the concentration of Si, which is assumed to be equal to the electron concentration), \textit{N$_{0}$} a normalizing impurity concentration, and $\alpha$ a constant with a value lower than unity. As reported in Ref.~\cite{Hilsum1974}, the normalizing impurity concentration N$_{0}$ is 10$^{17}$~cm$^{-3}$ for a wide variety of semiconducting materials, including GaAs. Hence, to fit eq.~(\ref{Hilsum}) to our experimental data, we fix N$_{0}$ to 10$^{17}$~cm$^{-3}$ and use $\mu_{L}$ and $\alpha$ as fitting parameters. This fit yields $\mu_{L}=(8600\pm600)$~cm$^{2}$/(V$\cdot$s) and $\alpha=0.43\pm0.03$. The obtanied value of $\mu_{L}$ is not far from those reported in Refs.~\cite{Kamp1994,Hilsum1974} for Si-doped GaAs layers, namely, 10000~cm$^{2}$/V$\cdot$s. Regarding $\alpha$, the obtained value is close to 0.5, i.\,e., to the characteristic value found by Hilsum for different n-type semiconducting materials \cite{Hilsum1974}. Our value of $\alpha$ is furthermore close to 0.4, the value reported in Refs.~\cite{Kamp1994,Hilsum1974} for Si doped GaAs. Based on the above results, we find that Si ionized impurities are responsible for the mobility reduction at high electron densities.

\subsection{GaAs doping using H$_{2}$-diluted CBr$_{4}$}

\begin{figure*}
\includegraphics*[width=1\textwidth]{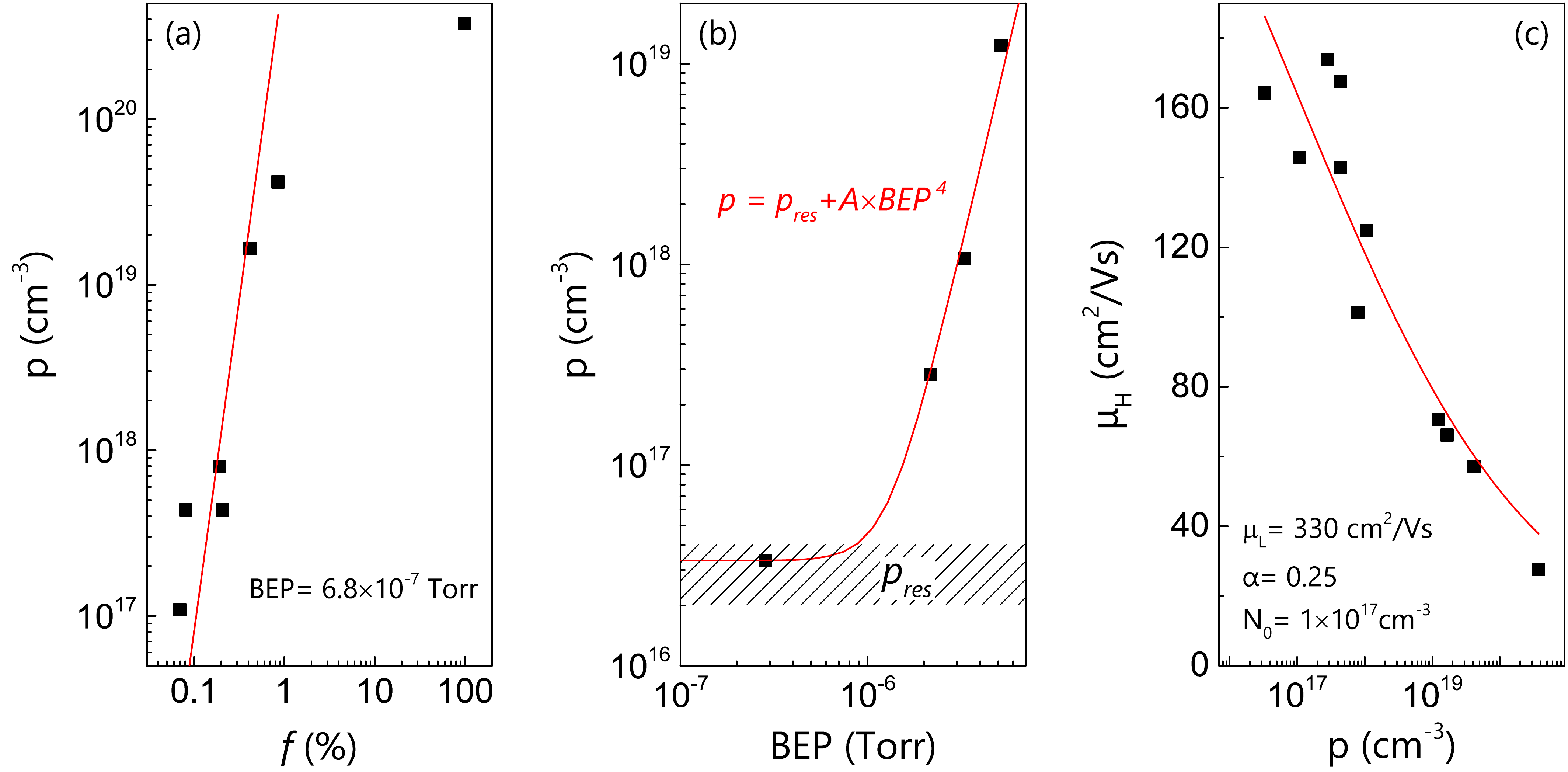}
\caption{\label{CBr4} Hole concentration, as determined by Hall effect measurements, as a function of (a) the dilution factor and (b) the BEP of the H$_{2}$-diluted CBr$_{4}$ precursor. The BEP in (a) is $6.8\times10^{-7}$~Torr and the dilution factor in (b) is $0.08\%$. The solid lines in (a) and (b) are fits of a fourth order power law to the experimental data assuming a residual hole concentration $p_{res}$ of $3\times10^{16}$~cm$^{-3}$. The only fitting parameter is thus the power prefactor \textit{ A}. (c) Hall electron mobility dependence on the hole concentration. The solid line is a fit of eq.(\ref{Hilsum}) to the experimental data.}
\end{figure*}

We analyze next the electrical properties of GaAs layers doped with C using H$_{2}$-diluted CBr$_{4}$ as gaseous precursor. Figure~\ref{CBr4}(a) shows the hole concentration, \textit{p}, as a function of the CBr$_{4}$ dilution factor, \textit{f}. The BEP of H$_{2}$-diluted CBr$_{4}$ is the same for all samples, 6.8$\times10^{-7}$~Torr. Despite of the slight spreading of the experimental data, attributed again to the partial condensation of CBr$_{4}$ inside the reservoir during the preparation of the different gas mixtures (as in the case of DTBSi), we observe a clear superlinear increase of \textit{p} with increasing values of \textit{f}. Specifically, \textit{p} increases from 1$\times10^{17}$ to 4$\times10^{19}$~cm$^{-3}$ as \textit{f} is varied between 0.07 and 1~$\%$. The maximum hole concentration achieved here, 3.8$\times10^{20}$~cm$^{-3}$ for a \textit{f} value of $100~\%$, compares well with the values reported in Refs.~\cite{Lyon1991,Houng1993,Westwater1997} for GaAs layers doped with C using pure CBr$_{4}$ as gas precursor. In contrast, to the best of our knowledge, hole concentrations in the 10$^{17}$~cm$^{-3}$ range have never been reported using CBr$_{4}$. The dilution of CBr$_{4}$ in H$_{2}$ is thus an effective approach for reaching low C concentrations in a controlled fashion.

As demonstrated above for H$_{2}$-diluted DTBSi, the C doping level can also be tuned by varying the BEP of the precursor while keeping constant the CBr$_{4}$ dilution factor. Figure~\ref{CBr4}(b) shows the variation of \textit{p} with the BEP of H$_{2}$-diluted CBr$_{4}$ for a \textit{f} value of 0.08~$\%$. The hole concentration monotonically increases from 3$\times10^{16}$~cm$^{-3}$ (a value that lays within the range of the typical residual C concentrations obtained in our chamber for non-intentionally doped GaAs layers, from 2 to  $4\times10^{16}$~cm$^{-3}$) to 1.2$\times10^{19}$~cm$^{-3}$ as the BEP is varied between $2.8\times10^{-7}$ and 5.2$\times10^{-6}$~Torr. As illustrated by the fit shown in the Fig.~\ref{CBr4}(b), the dependence of \textit{p} on the BEP is well described by a power law with an exponent of 4. Such a power law also properly describes the data presented in Fig.~\ref{CBr4}(a) for doping concentrations below the maximum value. Interestingly, this result is in contrast to the first order power law dependence on the CBr$_{4}$ mole fraction found by Buchan \textit{et al.} for the C doping of GaAs layers grown by MOVPE using TMGa and AsH$_{3}$ as Ga and As precursors, respectively \cite{Buchan1991}. We tentatively attribute this discrepancy to the inherent differences that characterize CBE and MOVPE growth tecnhniques, i.\,e., chemical reactions at the sample surface in CBE versus chemical reactions in the gas phase in MOVPE. In addition, the results obtained here could also differ from previous ones due to the different Ga precursor used in this work, TEGa. Notice that, as pointed out in Ref.~\cite{Buchan1991}, CBr$_{4}$ could react in concert with the Ga precursor or its reaction products to incorporate C into GaAs. 

The dependence of the hole mobility on the doping level, as determined by Hall effect, is summarized in Fig.~\ref{CBr4}(c). The mobility decreases from about 175 to 28~cm$^{2}$/V$\cdot$s as the hole concentration increases from 3$\times10^{16}$ to 4$\times10^{20}$~cm$^{-3}$. The obtained hole-cocentration dependent mobilities are comparable to those reported for C-doped GaAs layers grown by either MOVPE or MOMBE using different types of precursors, including the by-products of the chemical reactions occurring between TMGa and AsH$_{3}$, CCl$_{4}$, or pure CBr$_{4}$ \cite{Wang1991,Kim1993,Richter1995,Son1996}. As previously obtained for the Si-doped layers, the decreasing mobility with increasing doping level is properly explained considering the scattering caused by ionized C impurities. The best fit of eq.~(\ref{Hilsum}) to our experimental data with N$_{0}$ equal to 1$\times10^{17}$~cm$^{-3}$ yields $\mu_{L}$ and $\alpha$ values of $(330\pm20)$~cm$^{2}$/V$\cdot$s and $0.25\pm0.04$, respectively. These values are similar to those reported in Refs.~\cite{Kim1993,Richter1995,Son1996} for C-doped GaAs, i.\,e., N$_{0}=3\times10^{17}$~cm$^{-3}$ and $\alpha=0.3$. In view of these results, we conclude that the use of H$_{2}$ to dilute CBr$_{4}$ has no detrimental effect on the hole mobility, which is exclusively determined by the actual C concentration. 

\subsection{Luminescence from GaAs layers doped with H$_{2}$-diluted DTBSi and CBr$_{4}$}
\begin{figure*}
\centering
\includegraphics*[width=0.85\textwidth]{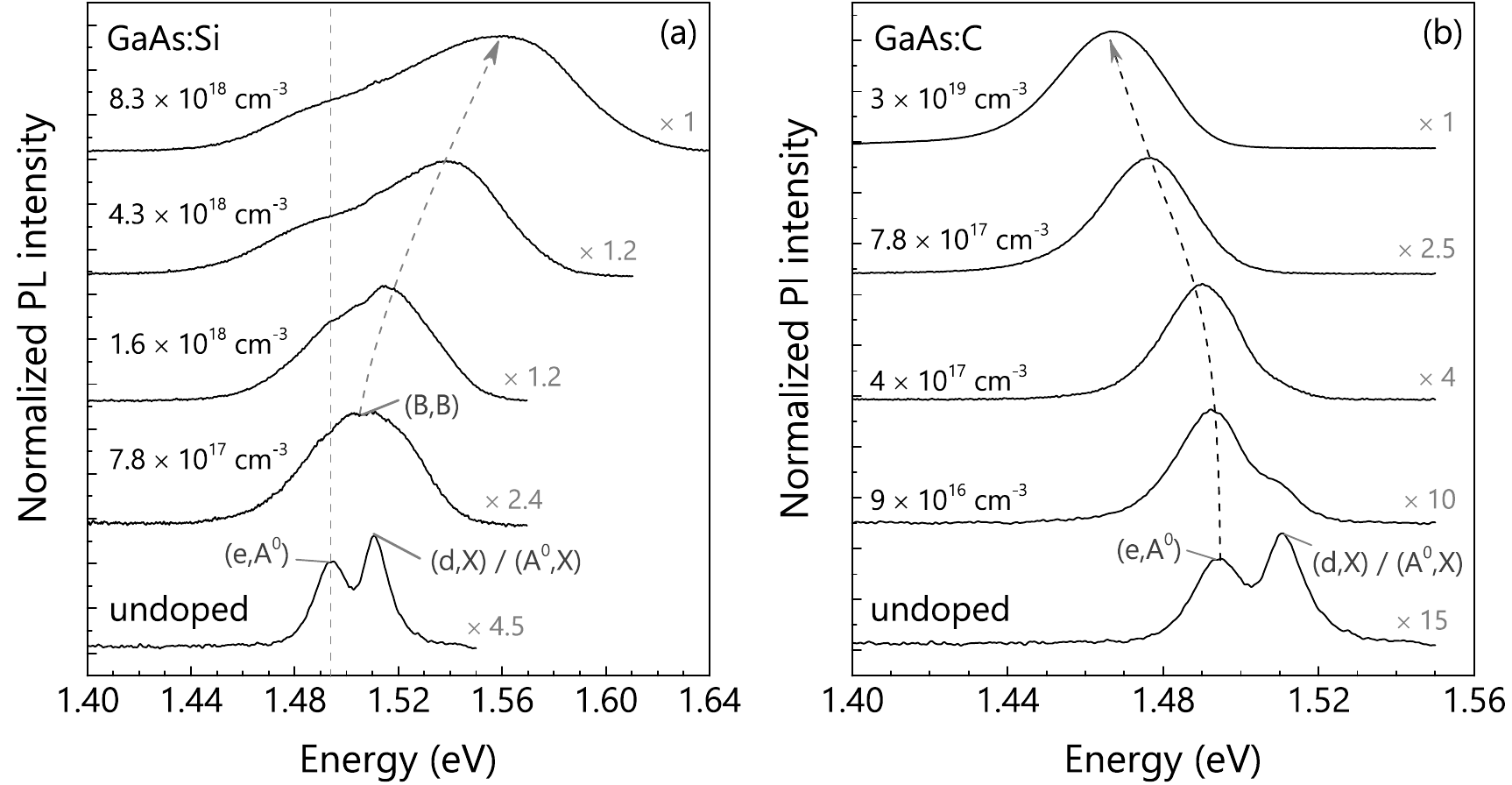}
\caption{\label{PL} (Color online) Normalized to the maximum intensity and vertically-shifted low-temperature (12~K) PL spectra of GaAs layers doped with different concentrations of (a) Si and (b) C using, respectively, H$_{2}$-diluted DTBSi and CBr$_{4}$ as gas precursors. In (a) and (b), we also include the spectrum from a non-intentionally doped GaAs layer with a residual C concentration of $3\times10^{16}$~cm$^{-3}$. The normalization intensity factors are indicated at the right side of each spectrum.}
\end{figure*}

We examine the optical properties of the GaAs layers doped with H$_{2}$-diluted DTBSi or CBr$_{4}$ by low-temperature (12 K) PL. Figure~\ref{PL}(a) shows the PL spectra of GaAs layers doped with different Si concentrations by using H$_{2}$-diluted DTBSi. We also include the PL spectrum of a non-intentionally doped GaAs layer grown in the same system, which exhibits a residual C concentration of $3\times10^{16}$~cm$^{-3}$. For clarity purposes, the spectra are normalized to the maximum intensity and vertically shifted. The spectrum of the reference sample has two different peaks centered at approximately 1.494 and 1.511~eV. The peak at 1.494~eV is attributed to a band-to-acceptor (e,A$^{0}$) transition, caused by the residual incorporation of C, and the one at 1.511~eV to excitons bound to neutral acceptors (A$^{0}$,X) and/or (d,X) neutral point defects \cite{Park1994,Pavesi1994,Briones1982}. For the Si-doped samples, we observe the (e,A$^{0}$) transition together with a second transition at a higher energies. This second transition, assigned to band-to-band recombination (B,B) \cite{Lideikis1989}, blue shifts from 1.51 to 1.56~eV and broadens as the Si concentration increases from $7.8\times10^{17}$ to $8.3\times10^{18}$~cm$^{-3}$. Since GaAs becomes degenerated at 12~K for Si concentrations above $2\times10^{18}$~cm$^{-3}$, the broadening and blue-shift of the (B,B) transition is explained in terms of the Burstein-Moss effect \cite{Lideikis1989}. Such an evolution of the (B,B) transition with increasing Si doping level is characteristic of GaAs:Si layers regardless of the growth technique \cite{Lideikis1989,Pavesi1997}.

Figure~\ref{PL}(b) presents the normalized to the maximum intensity and vertically-shifted PL spectra of GaAs layers doped with different levels of C using H$_{2}$-diluted CBr$_{4}$. For comparison, we present here again the spectrum of the non-intentionally doped GaAs layer. In accordance to Refs.\cite{Ozeki1974,Chen1994,Feng1995}, the PL spectra of the C-doped layers are dominated by the (e,A$^{0}$) transition. This transition red shifts and broadens with increasing C concentration. Specifically, the peak energy shifts from 1.494 to 1.467~eV when increasing the C concentration from the residual value up to $3\times10^{19}$~cm$^{-3}$. The red-shift of the (e,A$^{0}$) transition is due to the band gap narrowing \cite{Wang1994,Feng1995,Lee1996}, while the increasing linewidth is caused by the Burstein-Moss effect and the relaxation of the k-selection rules \cite{Wang1994}. Both the red-shift and the broadening of the peak are typical of highly C doped samples \cite{Chen1994,Wang1994,Feng1995,Lee1996}. Therefore, as in the case of the Si-doped GaAs layers doped with H$_{2}$-diluted DTBSi, the luminescence properties of the GaAs doped with H$_{2}$-diluted CBr$_{4}$ are comparable to those of equivalent samples prepared by other techniques using different dopant sources. Consequently, the use of H$_{2}$ to dilute either DTBSi or CBr$_{4}$ does not have any particular effect on the optical properties of Si and C doped GaAs layers grown by CBE. It is also important to highlight the observed increase on the PL intensity when increasing either the Si or C doping level; this general trend may be explained on the basis of the larger available electron concentration on Si-doped samples, and by an increasing number of recombination centers on C-doped samples.\\ 

\section{Summary and conclusions}
We have demonstrated the n- and p-type doping of GaAs layers in CBE using H$_{2}$-diluted DTBSi and CBr$_{4}$ as gas precursors. The dilution of pure DTBSi and CBr$_{4}$ in H$_{2}$ allowed us to achieve a wide range of doping levels. Specifically, GaAs electron and hole concentrations were varied from $6\times10^{17}$ to $1.2\times10^{19}$~cm$^{-3}$ and from $9\times10^{16}$ to $3.7\times10^{20}$~cm$^{-3}$, respectively. In the case of H$_{2}$-diluted DTBSi, the incorporation of Si into GaAs is found to follow a first order power law with respect to the dopant flux. Hence, this precursor is cracked in the cell providing active SiH$_{x}$ radicals to the sample surface. This first order power law is analogous to the reported dependence of the Si concentration on the partial pressure of pure DTBSi in MOVPE. In contrast, the dependence of the C incorporation on the flux of uncracked H$_{2}$-diluted CBr$_{4}$ follows not a first but a fourth order power law, revealing the occurrence of complex reactions at the sample surface. According to our results, for a given dilution factor of either DTBSi or CBr$_{4}$ in H$_{2}$, it is possible to obtain an accurate control on the doping level by varying the flux of the H$_{2}$-diluted precursor. However, we did not achieve a proper degree of control when varying the precursor dilution factor while keeping constant the flux. This result is attributed to the unintentional partial condensation of the precursor inside the reservoir during the dilution process, which results in a poor control on the actual precursor dilution factor. Consequently, a better degree of control is expected by improving the dilution process, such as avoiding condensation by heating the reservoir above the temperature of the pure precursor container. Finally, the analysis of the samples by Hall effect measurements and low-temperature PL do not indicate any detrimental or particular effect associated to H$_{2}$. Therefore, the use of H$_{2}$-diluted precursors allows to tune the doping level over a wide range without affecting the electrical and optical properties of GaAs. We expect that this approach to widen the range of achievable doping levels could be extended to other precursors and material systems.

\section{Acknowledgments}
This work was supported by Ministerio de Ciencia, Innovación y Universidades under Project No. TEC2016-78433-R. S. Fernández-Garrido and N. López acknowledge the final support received through the  Spanish  program  Ramón  y  Cajal  (co-financed  by  the  European  Social  Fund)  under  Grants  No. RYC-2016-19509 and RYC-2016-20588, respectively, from the former Ministerio de Ciencia, Innovación y Universidades. N. López  also acknowledges the funding received through the European ERC Starting Grant number 758885. 

\bibliographystyle{unsrt}
\bibliography{bibliography}

\end{document}